\documentstyle[mprocl]{article}

\input{psfig}
\def\be{\begin{equation}}
\def\ee{\end{equation}}
\def\bea{\begin{eqnarray}}
\def\eea{\end{eqnarray}}

\begin{document}

\title{STATIC REGULAR AND BLACK HOLE SOLUTIONS WITH AXIAL SYMMETRY
IN EYM AND EYMD THEORY}

\author{B. KLEIHAUS~\footnote{Speaker}, J. KUNZ}

\address{Fachbereich Physik, Universit\"at Oldenburg, Postfach 2503\\
D-26111 Oldenburg, Germany}

\maketitle\abstracts{
We discuss the recently discovered
new class of globally regular and black hole solutions
in Einstein-Yang-Mills and Einstein-Yang-Mills-dilaton theory.
These asymptotically flat solutions are static and 
possess only axial symmetry.
The black hole solutions possess a regular event horizon.}

SU(2) Einstein-Yang-Mills (EYM) 
and Einstein-Yang-Mills-dilaton (EYMD) theory possess
static spherically symmetric globally regular 
and black hole solutions.
These solutions are asymptotically flat
and possess non-trivial magnetic gauge field configurations,
but carry no global charge.
Being unstable, the globally regular solutions are
interpreted as gravitating sphalerons, while the black hole solutions
are considered as black holes inside sphalerons.
The black hole solutions
represent counterexamples to the ``no-hair'' conjecture.

In flat space spherically symmetric sphalerons
represent a special case of axially symmetric (multi)sphalerons \cite{kk}.
By constructing axially symmetric (multi)sphalerons in EYM and EYMD
theory, we here show, that the same holds true in curved space \cite{kk2}.
Furthermore, we demonstrate that the corresponding black hole solutions
with axial symmetry also exist \cite{kk3}.

The new class of solutions is based on the SU(2) EYMD action
\begin{equation}
S=\int \left ( \frac{R}{16\pi G} + L_M \right ) \sqrt{-g} d^4x
\   \end{equation}
with
\begin{equation}
L_M=-\frac{1}{2}\partial_\mu \Phi \partial^\mu \Phi
 -e^{2 \kappa \Phi }\frac{1}{2} {\rm Tr} (F_{\mu\nu} F^{\mu\nu})
\ , \end{equation}
$F_{\mu \nu} = 
\partial_\mu A_\nu -\partial_\nu A_\mu + i e \left[A_\mu , A_\nu \right] $,
and the Yang-Mills and dilaton coupling constants $e$ and $\kappa$.

The static axially symmetric ans\"atze for the metric
and the matter fields of the solutions are parametrized 
in terms of the coordinates $r$ and $\theta$, being
\begin{equation}
ds^2=
  - f dt^2 +  \frac{m}{f} d r^2 + \frac{m r^2}{f} d \theta^2 
           +  \frac{l r^2 \sin^2 \theta}{f} d\phi^2
\ , \label{metric} \end{equation}
and
\begin{equation}
A_\mu dx^\mu =
\frac{1}{2er} \left[ \tau^n_\phi 
 \left( H_1 dr + \left(1-H_2\right) r d\theta \right)
 -n \left( \tau^n_r H_3 + \tau^n_\theta \left(1-H_4\right) \right)
  r \sin \theta d\phi \right]
\ , \label{gf1} \end{equation}
with the Pauli matrices $\vec \tau = ( \tau_x, \tau_y, \tau_z) $ and
$\tau^n_r = \vec \tau \cdot 
(\sin \theta \cos n \phi, \sin \theta \sin n \phi, \cos \theta)$,
$\tau^n_\theta = \vec \tau \cdot 
(\cos \theta \cos n \phi, \cos \theta \sin n \phi, -\sin \theta)$,
$\tau^n_\phi = \vec \tau \cdot (-\sin n \phi, \cos n \phi,0)$.
We refer to $n$ as the winding number of the solutions.
The metric functions $f$, $m$ and $l$,
the gauge field functions $H_i$
and the dilaton function $\Phi$ depend only on $r$ and $\theta$.
For $n=1$ the spherically symmetric ansatz is recovered.
The system possesses a residual abelian gauge invariance
with respect to the transformation
$ U= e^{i\Gamma(r,\theta) \tau^n_\phi} $.
We choose the gauge condition
$ r \partial_r H_1 - \partial_\theta H_2 = 0 $.

The boundary conditions at infinity and along the $\rho$- and $z$-axis 
agree for the globally regular and the black hole solutions.
For asymptotically flat and magnetically neutral solutions
the boundary conditions at infinity are
\begin{equation}
f=m=l=1 \ , \ \ \ \Phi=0 \ , \ \ \
H_2=H_4=\pm 1, \ \ \ H_1=H_3=0
\ . \label{bc2} \end{equation}
Along the axes axial and parity reflection symmetry of the solutions
yield the boundary conditions
\begin{equation}
\partial_\theta f=\partial_\theta m=
\partial_\theta l=\partial_\theta \Phi=
\partial_\theta H_1=\partial_\theta H_3=0 \ , \ \ \
H_1=H_3=0
\ .  \end{equation}
For the globally regular solutions
the boundary conditions at the origin are
\begin{equation}
\partial_r f= \partial_r m= \partial_r l= \partial_r \Phi =0 \ , \ \ \
H_2=H_4= 1, \ \ \ H_1=H_3=0
\ . \end{equation}
The event horizon of the static black hole solutions
is characterized by $g_{tt}=-f=0$.
Imposing that the horizon of the black hole solutions
resides at a surface of constant $r$,
the boundary conditions at a regular horizon are
\begin{eqnarray}
&f=m=l=0 \ , \ \ \  \partial_r \Phi=0 \ ,  \ \ \
 \partial_\theta H_1 + r \partial_r H_2 = 0 \ ,
\nonumber \\
& r \partial_r H_3-H_1 H_4=0,   \ \ \ 
 r \partial_r H_4+H_1( H_3 + {\rm ctg} \theta) =0
\ . \label{bc3} \end{eqnarray}
The equations of motion yield only three boundary conditions
for the gauge field functions $H_i$;
the fourth condition is needed to fix the gauge at the horizon,
e.g.~by choosing $\partial_r H_1 = 0$.

\begin{figure}[h]
\rule{5cm}{0.2mm}\hfill\rule{5cm}{0.2mm}
\psfig{figure=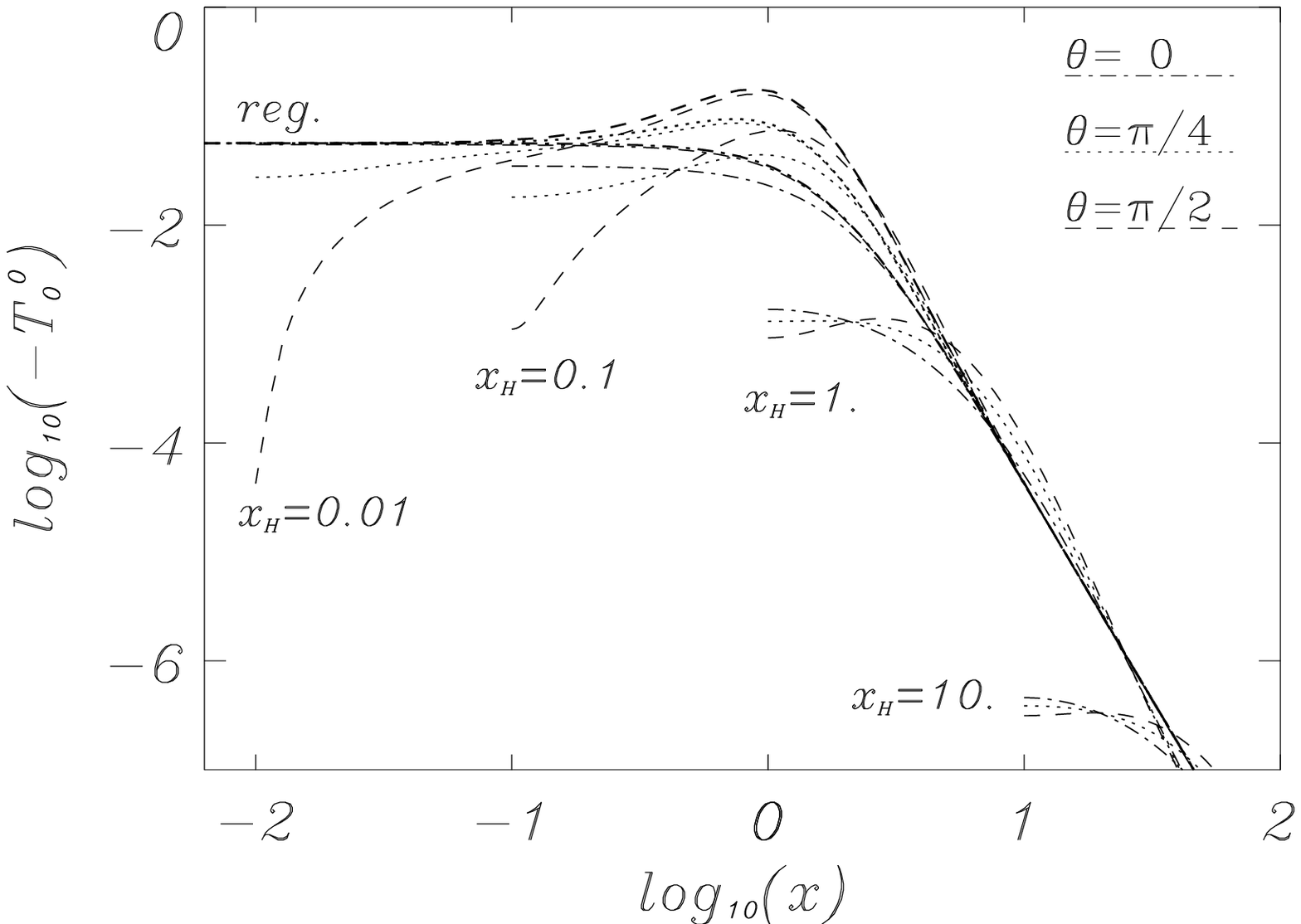,height=1.4in}
\psfig{figure=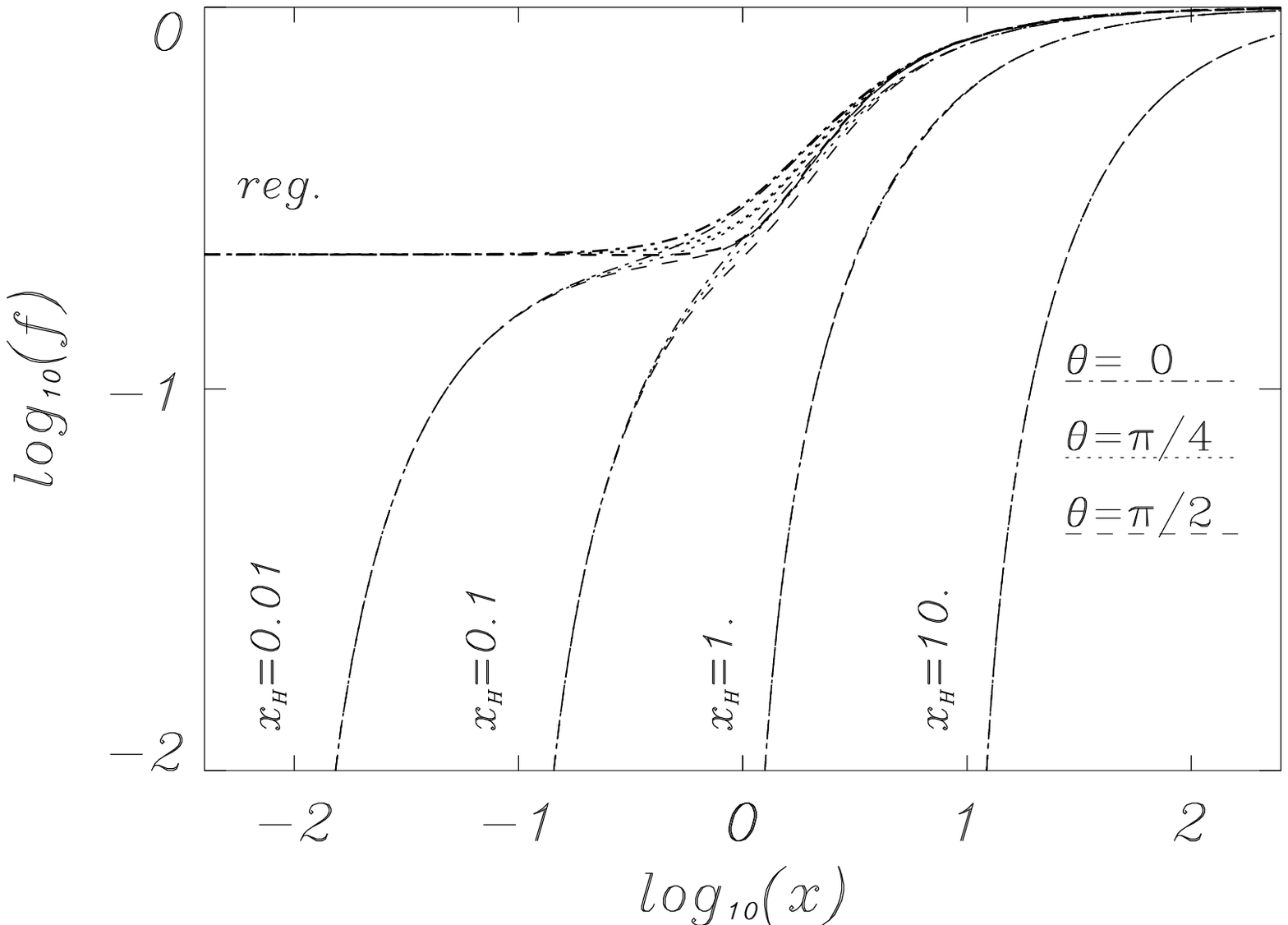,height=1.4in}
\rule{5cm}{0.2mm}\hfill\rule{5cm}{0.2mm}
\caption{
The energy density of the matter fields $\epsilon=-T_0^0$ 
and the metric function $f$ are shown as 
a function of the dimensionless coordinate $x$
for the angles $\theta=0$, $\pi/4$ and $\pi/2$
for several EYM black hole solutions 
as well as for the corresponding globally regular solution.
\label{fig:fig1}}
\end{figure}
The solutions depend on the winding number $n$, the node number $k$
of the gauge field functions, the dilaton coupling
constant $\gamma =\kappa/\sqrt{4\pi G}$
($\gamma=1$ corresponds to string theory) and the horizon radius 
$x_{\rm H}=(e/\sqrt{4\pi G}) r_{\rm H}$.
In Figs.~1-2 we show the globally regular solution 
and several black hole solutions for $n=2$, $k=1$, $\gamma=1$.

The energy density of the black hole solutions is angle-dependent
at the horizon.
We observe a strong peak of the energy density
on the $\rho$-axis (away from the horizon)
for the globally regular solution and for the black hole solutions with 
small $x_{\rm H}$. 
For larger values of $x_{\rm H}$ the global maximum 
occurs on the $z$-axis at the horizon.
With decreasing $x_{\rm H}$ 
the black hole solutions tend to the globally regular solutions.
The limit $x_{\rm H} \rightarrow 0$ is not smooth, however.
With increasing $x_{\rm H}$
the energy density of the matter fields becomes less important.
Therefore the metric functions become more spherical.

\begin{figure}[h]
\rule{5cm}{0.2mm}\hfill\rule{5cm}{0.2mm}
\psfig{figure=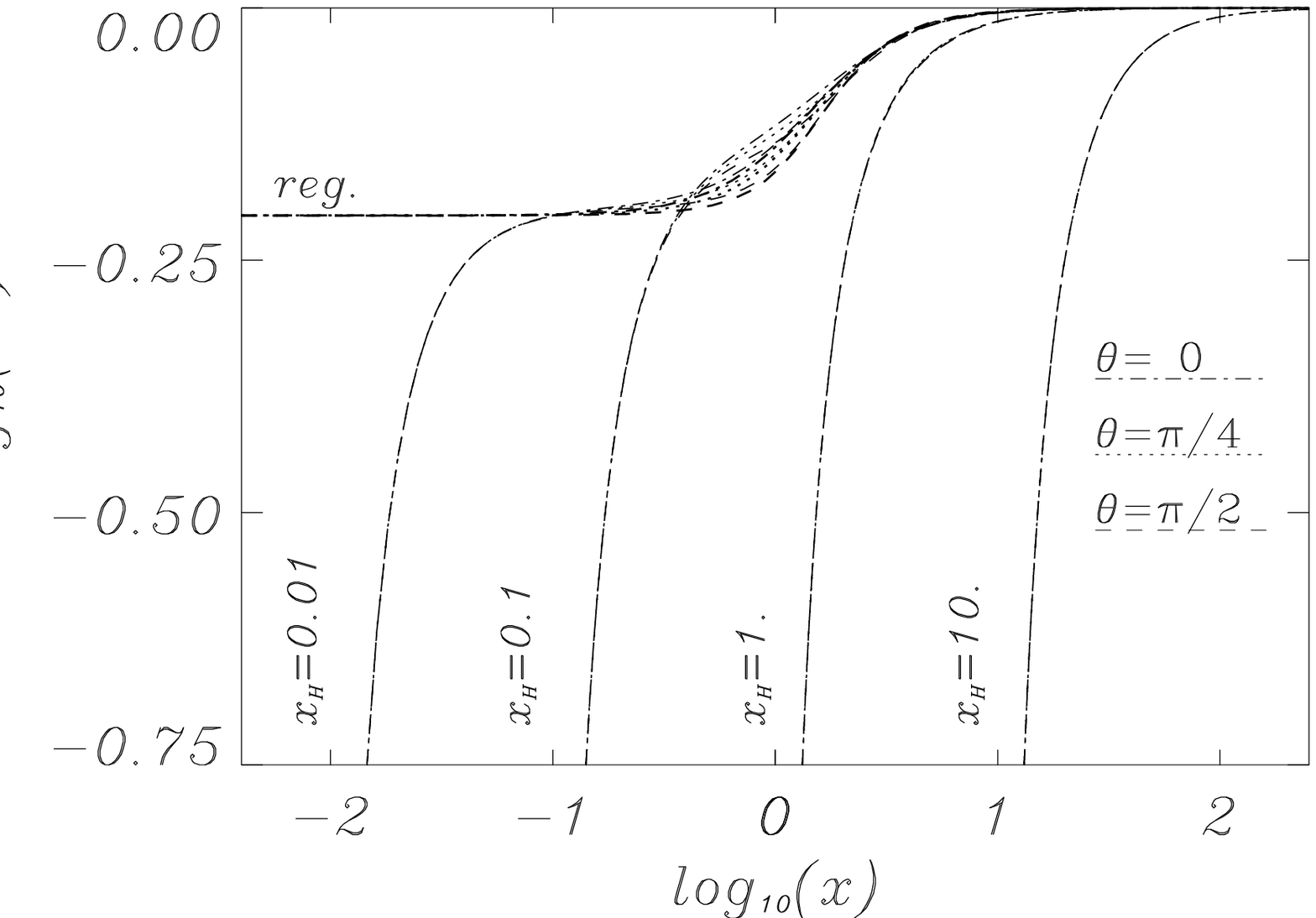,height=1.4in}
\psfig{figure=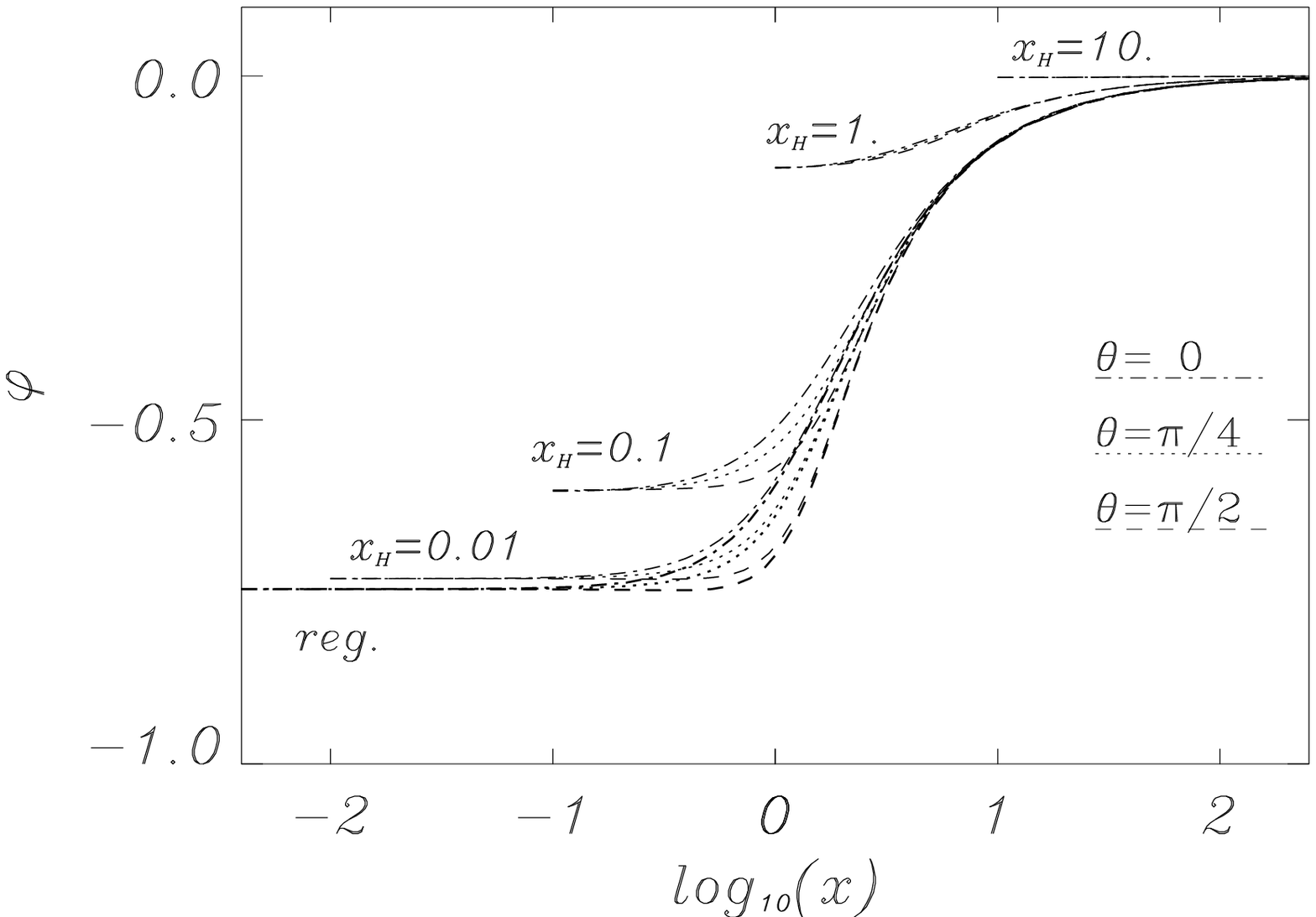,height=1.4in}
\rule{5cm}{0.2mm}\hfill\rule{5cm}{0.2mm}
\caption{
Same as Fig.~\ref{fig:fig1} for the metric function $m$ and the
dilaton function $\varphi$.
\label{fig:fig2}}
\end{figure}
For fixed winding number $n$ with increasing node number $k$
the solutions form sequences.
The sequences of magnetically neutral non-abelian
axially symmetric solutions with winding number $n$
tend to magnetically charged abelian spherically symmetric
limiting solutions, corresponding to
Einstein-Maxwell-dilaton solutions
for finite values of $\gamma$ and to Reissner-Nordstr\o m
solutions for $\gamma=0$, which carry $n$ units of magnetic charge.

The mass $M$ of the solutions can be obtained directly from
the total energy-momentum ``tensor'' $\tau^{\mu\nu}$
of matter and gravitation, $M=\int \tau^{00} d^3r$,
leading to the dimensionless mass $\mu =(e/\sqrt{4\pi G}) G M
 = \frac{1}{2} x^2 \partial_x f |_\infty $.
Similarly, the dilaton charge is given by 
$D  =  x^2 \partial_x \varphi |_\infty $.
The globally regular solutions satisfy the relation $D=\gamma \mu$.
At the horizon of the black hole solutions 
the Kretschmann scalar $R^{\mu\nu\alpha\beta}R_{\mu\nu\alpha\beta}$
is finite and the surface gravity $\kappa_{\rm sg}$ is constant.
The black hole solutions satisfy the relation $D=\gamma (\mu - 2 TS)$,
where the temperature $T$ is given by $T=\kappa_{\rm sg} /(2 \pi)$,
and the entropy $S$ is proportional to the area $A$ of the horizon, $S=A/4$.

\end{document}